\documentstyle[prd,aps,epsf,eqsecnum,twocolumn]{revtex}

\bibstyle{unsrt}
\begin{document}
\draft

\twocolumn[\hsize\textwidth\columnwidth\hsize\csname
@twocolumnfalse\endcsname

\title{Complex Square Well --- A New Exactly Solvable Quantum Mechanical Model}

\author{Carl M. Bender$^1$, Stefan Boettcher$^2$, H. F. Jones$^3$,
and Van M. Savage$^1$}
\address{${}^1$Department of Physics, Washington University, St. Louis, MO
63130, USA}
\address{${}^2$Department of Physics, Emory University, Atlanta, GA 30322, USA}
\address{${}^3$Blackett Laboratory, Imperial College, London SW7 2BZ, UK}

\date{\today}

\maketitle
 
\begin{abstract}
Recently, a class of ${\cal PT}$-invariant quantum mechanical models described
by the non-Hermitian Hamiltonian $H=p^2+x^2(ix)^\epsilon$ was studied. It was
found that the energy levels for this theory are real for all $\epsilon\geq0$.
Here, the limit as $\epsilon\to\infty$ is examined. It is shown that in this
limit, the theory becomes exactly solvable. A generalization of this
Hamiltonian, $H=p^2+x^{2M}(ix)^\epsilon$ ($M=1,2,3,\ldots$) is also studied, and
this ${\cal PT}$-symmetric Hamiltonian becomes exactly solvable in the
large-$\epsilon$ limit as well. In effect, what is obtained in each case is a
complex analog of the Hamiltonian for the square well potential. Expansions
about the large-$\epsilon$ limit are obtained.
\end{abstract}

\pacs{11.30.Er, 3.65.-w, 11.10.Jj, 11.25.Db}
]
 
\section{Introduction}
\label{s1}

The infinite square-well potential,
\begin{equation}
V_{\rm SW}(x)=\left\{
\begin{array}{cl}
0&(|x|<1),\\ 1&(|x|=1),\\ \infty&(|x|>1),
\end{array}\right.
\label{eq1.1}
\end{equation}
is the simplest of all quantum potentials. It is studied at the beginning of any
introductory class in quantum mechanics. This model is a useful teaching tool
because the eigenvalues and eigenfunctions for this potential can all be found
in closed form.

The infinite square-well potential can be regarded as the limiting case of a
class of potentials of the form
\begin{equation}
V_M(x)=x^{2M}\qquad(M=1,2,3,4,\ldots).
\label{eq1.2}
\end{equation}
Here, as $M\to\infty$, $V_M(x)\to V_{\rm SW}(x)$.

The eigenvalues of the Hamiltonian for $V_M$,
\begin{equation}
H=p^2+x^{2M},
\label{eq1.3}
\end{equation}
can only be found in closed form for the special case of the harmonic oscillator
$M=1$. For all other positive integer values of $M$ there is no exact solution
to these anharmonic oscillators. Thus, the only two exactly solvable cases known
are the extreme lower and upper limits $M=1$ and $M\to\infty$. The asymptotic
behavior of the eigenvalues of $H$ in Eq.~(\ref{eq1.3}) for large $M$ was
studied in Ref.~\cite{r1}.

\begin{figure}
\epsfxsize=2.2truein
\hskip 0.15truein\epsffile{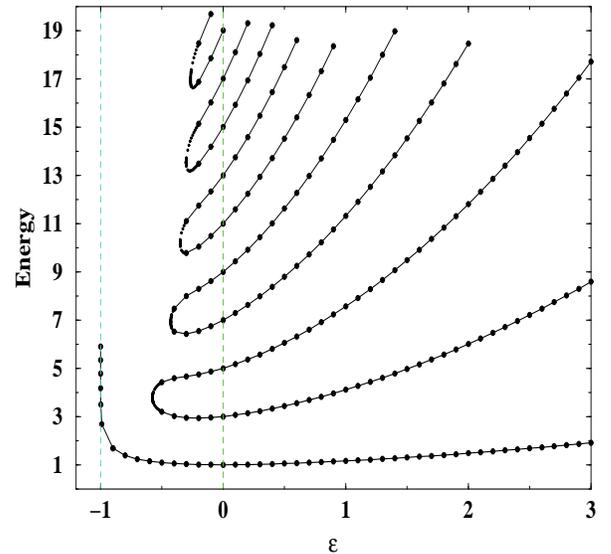}
\caption{
\narrowtext
Energy levels of the Hamiltonian $H=p^2+x^2(ix)^\epsilon$ as functions of the
parameter $\epsilon$. There are three regions: When $\epsilon\geq0$ the spectrum
is entirely real and positive. All eigenvalues rise monotonically with
increasing $\epsilon$. The lower bound of this region, $\epsilon=0$, corresponds
to the harmonic oscillator, whose energy levels are $E_k=2k+1$. When $-1<
\epsilon<0$, there are a finite number of real positive eigenvalues and an
infinite number of complex conjugate pairs of eigenvalues. As $\epsilon$
decreases from $0$ to $-1$, the number of real eigenvalues decreases. As
$\epsilon$ approaches $-1^+$, the ground-state energy diverges. For $\epsilon
\leq-1$ there are no real eigenvalues.}
\label{fig1}
\end{figure}

In a recent letter \cite{r2} the spectra of the class of non-Hermitian
${\cal PT}$-symmetric Hamiltonians of the form 
\begin{equation}
H=p^2+x^2(ix)^\epsilon\qquad(\epsilon\geq0)
\label{eq1.4}
\end{equation}
were shown to be real and positive. It is believed that the reality and
positivity of the spectra are a consequence of ${\cal PT}$ symmetry. Here, the
case $\epsilon=0$ is again the harmonic oscillator. For finite values of
$\epsilon$ larger than $0$ there is no exact analytical solution for the
eigenvalues. However, solutions can be found by numerical integration; the
eigenvalues of $H$ in Eq.~(\ref{eq1.4}) as functions of $\epsilon$ are
displayed in Fig.~\ref{fig1}.

In the past we have always regarded the parameter $\epsilon$ as being small;
we have defined theories by analytically continuing away from $\epsilon=0$.
However, in this paper we investigate the {\it large}-$\epsilon$ limit of the
Hamiltonian in Eq.~(\ref{eq1.4}). We will show that in this limit the theory
becomes exactly solvable. An exact formula for the $k$th energy level in the
limit of large $\epsilon$ is
\begin{equation}
E_k(\epsilon)\sim{1\over4}\left(k+{1\over2}\right)^2\epsilon^2\qquad(\epsilon\to
\infty).
\label{eq1.5}
\end{equation}

More generally, we will consider the large-$\epsilon$ limit of an {\it infinite}
number of classes of ${\cal PT}$-symmetric Hamiltonians of the form \cite{r3}
\begin{equation}
H=p^2+x^{2M}(ix)^\epsilon\qquad(\epsilon\geq0,~M=1,2,3,\ldots).
\label{eq1.6}
\end{equation}
For each positive integer value of $M$, these Hamiltonians may be regarded as
complex deformations of the Hermitian Hamiltonian $H=p^2+x^{2M}$ in
Eq.~(\ref{eq1.3}). In the limit as $\epsilon\to\infty$ each of these
Hamiltonians becomes exactly solvable; the spectrum for large $\epsilon$ is
given by
\begin{equation}
E_k(M,\epsilon)\sim{1\over4}\left(k+{P\over M+1}\right)^2\epsilon^2\qquad
(\epsilon\to\infty),
\label{eq1.7}
\end{equation}
where $P=1,2,3,\ldots,M$.

For the Hamiltonian $H$ in Eq.~(\ref{eq1.6}) the Schr\"odinger differential
equation corresponding to the eigenvalue problem $H\psi=E\psi$ is 
\begin{equation}
-\psi''(x)+x^{2M}(ix)^\epsilon\psi(x)=E\psi(x).
\label{eq1.8}
\end{equation}
To obtain real eigenvalues from this equation it is necessary to define the
boundary conditions properly. The regions in the cut complex-$x$ plane in which
$\psi(x)$ vanishes exponentially as $|x|\to\infty$ are wedges. In
Refs.~\cite{r2,r3} the wedges for $\epsilon>0$ were chosen to be the analytic
continuations of the wedges for the anharmonic oscillator ($\epsilon=0$), which
are centered about the negative and positive real axes and have angular opening
$\pi/(M+1)$. This analytic continuation defines the boundary conditions in the
complex-$x$ plane. For arbitrary $\epsilon>0$ the anti-Stokes' lines at the
centers of the left and right wedges lie below the real axis at the angles
\begin{eqnarray}
\theta_{\rm left}&=&-\pi+{\epsilon\pi\over4M+2\epsilon+4},\nonumber\\
&&\nonumber\\
\theta_{\rm right}&=&-{\epsilon\pi\over4M+2\epsilon+4}.
\label{eq1.9}
\end{eqnarray}
The opening angle of each of  these wedges is $2\pi/(2M+\epsilon+2)$. In
Refs.~\cite{r2,r3} the time-independent Schr\"odinger equation was integrated
numerically inside the wedges to determine the eigenvalues to high precision.
Observe that as $\epsilon$ increases from its anharmonic oscillator value
($\epsilon=0$), the wedges bounding the integration path undergo a continuous
deformation as a function of $\epsilon$. As $\epsilon$ increases, the opening
angles of the wedges become smaller and both wedges rotate downward towards the
negative-imaginary axis. Also, note that the angular difference
$\theta_{\rm right}-\theta_{\rm left}=2\pi(M+1)/\epsilon$ approaches zero as
$\epsilon$ increases.

This paper is organized very simply. In Sec.~\ref{s2} we consider the special
case $M=1$ in Eq.~(\ref{eq1.4}). Then in Sec.~\ref{s3} we generalize to the case
of arbitrary integer $M$ in Eq.~(\ref{eq1.6}). Finally, in Sec.~\ref{s4} we
examine expansions about the $\epsilon\to\infty$ limit of the theory.

\section{Special Case $M=1$.}
\label{s2}

The eigenvalues of $H$ in Eq.~(\ref{eq1.4}) can be found approximately using WKB
theory. The left and right turning points for this calculation lie inside the
left and right wedges at
\begin{eqnarray}
x_{\rm left}&=&E^{1/(2+\epsilon)}\exp\left(-i\pi+{\epsilon\over2\epsilon+4}i\pi
\right),\nonumber\\
x_{\rm right}&=&E^{1/(2+\epsilon)}\exp\left(-{\epsilon\over2\epsilon+4}i\pi
\right).
\label{eq2.1}
\end{eqnarray}
As explained in Ref.~\cite{r2}, the WKB quantization formula is
\begin{equation}
\left(k+{1\over2}\right)\pi\sim\int_{x_{\rm left}}^{x_{\rm right}}dx\,\sqrt{E-
x^2(ix)^\epsilon}\quad(k\to\infty),
\label{eq2.2}
\end{equation}
where the path of integration is a curve from the left turning point to the
right turning point along which the quantity $dx$ times the integrand of
Eq.~(\ref{eq2.2}) is {\it real}. This path lies in the lower-half $x$ plane and
is symmetric with respect to the imaginary axis. The path resembles an inverted
parabola; it emerges from the left turning point and rises monotonically until
it crosses the imaginary axis; it then falls monotonically until it reaches the
right turning point. As calculated in Ref.~\cite{r2}, the WKB quantization
formula (\ref{eq2.2}) gives
\begin{equation}
E_k\sim\left[
{\Gamma\left({3\epsilon+8\over2\epsilon+4}\right)\sqrt{\pi}\left(k+{1\over2}
\right)\over\sin\left({\pi\over\epsilon+2}\right)\Gamma\left({\epsilon+3\over
\epsilon+2}\right)}\right]^{2\epsilon+4\over\epsilon+4}\quad(k\to\infty).
\label{eq2.3}
\end{equation}

When the parameter $\epsilon$ is large, the right side of Eq.~(\ref{eq2.3})
simplifies dramatically and we have the result in Eq.~(\ref{eq1.5}) with
corrections of order $\epsilon\ln\epsilon$. As we will see, this happens to be
the exact answer for all energy levels; that is, for {\it all} values of $k$.
Since the WKB formula in Eq.~(\ref{eq2.3}) is only valid for large $k$ with
$\epsilon$ fixed, it is not at all obvious why the leading-order WKB
calculation gives the exact answer.

It is surprising to learn that the energy levels grow as $\epsilon^2$ for large
$\epsilon$. Recall that the energy levels of the Hamiltonian in
Eq.~(\ref{eq1.3}) approach finite limits as $M\to\infty$. [These limits are the
energy levels of the conventional square well $V_{\rm SW}$ in
Eq.~(\ref{eq1.1}).] To understand why the energy levels for the
${\cal PT}$-symmetric Hamiltonian in Eq.~(\ref{eq1.4}) grow as $\epsilon^2$ we
use the uncertainty principle. From Eq.~(\ref{eq2.1}) we see that the turning
points rotate towards each other as $\epsilon\to\infty$. (They both approach the
point $-i$ on the negative imaginary axis.) Indeed, the distance between the
turning points is of order $1/\epsilon$. [For the case of the Hamiltonian
$H$ in Eq.~(\ref{eq1.3}) the turning points stabilize at $\pm1$ as $M\to
\infty$.] Thus, the quantum particle is trapped in a region whose size
$\Delta x$ is of order $1/\epsilon$. The uncertainty in the momentum $\Delta p$
of the particle is therefore of order $\epsilon$. Finally, since the energy is
the square of the momentum, we conclude that the energy levels must be of order
$\epsilon^2$.

Let us rederive this result using the time-energy version of the uncertainty
principle. As explained in Refs.~\cite{r2,r3}, a {\it classical} particle
described by the Hamiltonian in Eq.~(\ref{eq1.4}) exhibits periodic motion. The
period $T$ of this complex pendulum is given exactly by the formula
\begin{eqnarray}
T=4\sqrt{\pi}E^{-{\epsilon\over4+2\epsilon}}{\Gamma\left({3+\epsilon\over2+
\epsilon}\right)\cos\left({\epsilon\pi\over4+2\epsilon}\right)\over
\Gamma\left({4+\epsilon\over4+2\epsilon}\right)}.
\label{eq2.4}
\end{eqnarray}
For large $\epsilon$ we have
\begin{eqnarray}
T\sim4\pi/(\epsilon\sqrt{E})\qquad(\epsilon\to\infty).
\label{eq2.5}
\end{eqnarray}
Multiplying this equation by $E$ gives the product $ET$ on the left side, which,
by the uncertainty principle is of order 1. Thus, solving for $E$, we find again
that $E$ is of order $\epsilon^2$ for large $\epsilon$.

This last calculation illustrates an important difference between conventional
quantum theories and ${\cal PT}$-symmetric quantum theories. In a conventional
Hermitian theory both the classical periodic motion and the WKB path of
integration coincide; this {\it classically allowed} region lies on the real
axis between the turning points. For ${\cal PT}$-symmetric theories the WKB
contour and the classical path do not coincide. The classical periodic motion
follows a path joining the turning points that, like the WKB path, is symmetric
about the negative imaginary axis. However, unlike the WKB path, the classical
path moves {\it downward} rather than upward as it approaches the
negative-imaginary axis (see, for example, Fig.~2 of Ref.~\cite{r3}).

Having discussed this problem heuristically, we now give a precise calculation
of the spectrum in the limit of large $\epsilon$. We begin by substituting
\begin{equation}
x=\left(-i+{z\pi\over2+\epsilon}\right)E^{1\over2+\epsilon}
\label{eq2.6}
\end{equation}
into Eq.~(\ref{eq1.8}) with $M=1$. The resulting differential equation for large
$\epsilon$ is
\begin{eqnarray}
{d^2\over dz^2}\psi(z)+F\pi^2\left(1+e^{iz\pi}\right)\psi(z)=0,
\label{eq2.7}
\end{eqnarray}
where
\begin{eqnarray}
F=E/\epsilon^2
\label{eq2.8}
\end{eqnarray}
and we have used the identity $\lim_{\epsilon\to\infty}(1+x/\epsilon)^\epsilon=
e^x$.

The advantage of the differential equation (\ref{eq2.7}) is that it is
independent of $\epsilon$. [As such, this equation corresponds to the
$M$-independent Schr\"odinger equation for the square well that is obtained from
Eq.~(\ref{eq1.3}) in the limit of large $M$.] In the variable $z$ the turning
points at $z=-1$ and $z=1$ are fixed and well separated in the limit of large
$\epsilon$. The large-$\epsilon$ behavior of $E$ in Eq.~(\ref{eq1.5}) is already
evident in Eq.~(\ref{eq2.8}). Imposing the appropriate boundary conditions on
Eq.~(\ref{eq2.7}) gives eigenvalues $F$ that are clearly independent of
$\epsilon$. Thus, for large $\epsilon$ we see that $E$ grows like $\epsilon^2$.

Because there is no longer any small parameter in Eq.~(\ref{eq2.7}), this
equation cannot be solved approximately using a perturbative method such as WKB.
It is necessary to solve this equation exactly. Fortunately, we can solve it
exactly by making a simple substitution. The change of variable
\begin{eqnarray}
w=2\sqrt{F}e^{i\pi z/2}
\label{eq2.9}
\end{eqnarray}
converts Eq.~(\ref{eq2.7}) to a modified Bessel equation \cite{r4}:
\begin{eqnarray}
w^2{d^2\over dw^2}\psi(w)+w{d\over dw}\psi(w)-\left(w^2+\nu^2\right)\psi(w)=0,
\label{eq2.10}
\end{eqnarray}
where
\begin{eqnarray}
\nu=2\sqrt{F}.
\label{eq2.11}
\end{eqnarray}
The exact solution to this equation is a linear combination of modified
Bessel functions \cite{r4}:
\begin{eqnarray}
\psi(w)=C_1I_\nu(w)+C_2K_\nu(w),
\label{eq2.12}
\end{eqnarray}
where $C_1$ and $C_2$ are arbitrary constants. Thus, in terms of the $z$
variable we have
\begin{eqnarray}
\psi(z)=C_1I_\nu\left(\nu e^{i\pi z/2}\right)+C_2K_\nu\left(\nu e^{i\pi z/2}
\right).
\label{eq2.13}
\end{eqnarray}

\begin{figure}
\epsfxsize=2.9truein
\hskip 0.15truein\epsffile{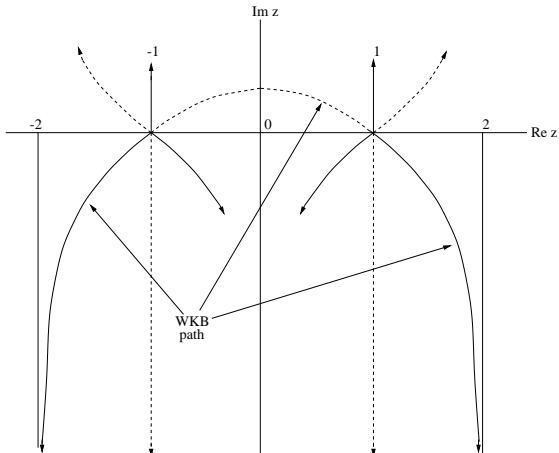}
\caption{
\narrowtext
Stokes' lines and anti-Stokes' lines for the differential equation
(\ref{eq2.7}). Three Stokes' lines (dashed lines) and three anti-Stokes'
lines (solid lines) emerge from the turning points at $z=\pm1$. The
path of integration for the WKB quantization condition in Eq.~(\ref{eq2.2})
corresponds to the arch-shaped dotted line connecting the turning points.}
\label{fig2}
\end{figure}

We must now impose boundary conditions on $\psi(z)$. Emanating from the turning
points at $z=-1$ and $z=1$ are three Stokes' lines (lines along which
the solution is purely oscillatory and not growing or falling exponentially) and
three anti-Stokes' lines (lines along which the solution is purely exponential
and not oscillatory). These Stokes' and anti-Stokes' lines are shown as dashed
and solid lines on Fig.~\ref{fig2}. The Stokes' lines emerge from the turning
points going up to the left and the right at $30^\circ$ and also directly
down. The anti-Stokes' lines emerge from the turning points going down to the
left and the right at $30^\circ$ and directly up. Note that the Stokes' line
going up to the right from the turning point at $z=-1$ joins continuously onto
the Stokes' line going up to the left from the turning point at $z=1$. The
anti-Stokes' lines going down to the left from $z=-1$ and down to the right from
$z=1$ eventually become vertical and asymptote to the lines ${\rm Re}\,z=-2$ and
${\rm Re}\,z=2$. We impose the boundary conditions that $\psi(z)\to0$ on these
anti-Stokes' lines because these correspond to the center lines of the wedges in
Eq.~(\ref{eq1.9}) in the complex-$x$ plane (for $M=1$).

To summarize, in the large-$\epsilon$ limit of the Hamiltonian in
Eq.~(\ref{eq1.4}), the eigenvalue problem for the scaled eigenvalues $F$ is a
two-turning-point problem that lies along an arch-shaped contour. The legs of
the arches lie below the real-$z$ axis and approach $\pm2-i\infty$. The turning
points at $z=\pm1$ are joined by the Stokes' line lying above the real-$z$ axis
as indicated in Fig.~\ref{fig2}. This is the complex version of the
infinite-square-well problem in elementary quantum mechanics. In the square-well
problem there are also two turning points at $\pm1$ joined by a Stokes' line
lying on the real axis. However, there are no anti-Stokes' lines along which the
wave function dies away exponentially; the wave function simply vanishes at the
turning points.

The quantized energy levels are determined by imposing the boundary conditions
discussed above on the modified Bessel functions in Eq.~(\ref{eq2.13}). For
simplicity, we impose these conditions on the vertical lines $z=\pm2-iy$, where
$y\to+\infty$. In terms of the variable $y$ the wave function $\psi$ in
Eq.~(\ref{eq2.13}) becomes
\begin{equation}
\psi(y)=C_1I_\nu\left(\nu e^{-i\pi}e^{\pi y/2}\right)+C_2K_\nu\left(\nu e^{-i
\pi}e^{\pi y/2}\right)
\label{eq2.14}
\end{equation}
at $z=-2-iy$, and
\begin{eqnarray}
\psi(y)=C_1I_\nu\left(\nu e^{i\pi}e^{\pi y/2}\right)
+C_2K_\nu\left(\nu e^{i\pi}e^{\pi y/2}\right)
\label{eq2.15}
\end{eqnarray}
at $z=2-iy$.

Our objective now is to simplify these equations by making the arguments of the
modified Bessel equations entirely real and positive. To do so we use the
following functional equations satisfied by $I_\nu$ and $K_\nu$ \cite{r4}:
\begin{eqnarray}
I_\nu\left(e^{m\pi i}z\right)&=&e^{m\nu\pi i}I_\nu(z),\nonumber\\
K_\nu\left(e^{m\pi i}z\right)&=&e^{-m\nu\pi i}K_\nu(z)
-i\pi{\sin(m\nu\pi)\over\sin(\nu\pi)}I_\nu(z),
\label{eq2.16}
\end{eqnarray}
where $m$ is an integer. According to these relations, Eq.~(\ref{eq2.14})
becomes
\begin{eqnarray}
\psi(y)&=&C_1e^{-\nu\pi i}I_\nu\left(\nu e^{\pi y/2}\right)\nonumber\\
&+&C_2\left[e^{\nu\pi i}K_\nu\left(\nu e^{\pi y/2}\right)+i\pi
I_\nu\left(\nu e^{\pi y/2}\right)\right]
\label{eq2.17}
\end{eqnarray}
and Eq.~(\ref{eq2.15}) becomes
\begin{eqnarray}
\psi(y)&=&C_1e^{\nu\pi i}I_\nu\left(\nu e^{\pi y/2}\right)\nonumber\\
&+&C_2\left[e^{-\nu\pi i}K_\nu\left(\nu e^{\pi y/2}\right)-i\pi
I_\nu\left(\nu e^{\pi y/2}\right)\right].
\label{eq2.18}
\end{eqnarray}

Next, we use the asymptotic behavior of the modified Bessel functions for
large positive argument. The function $I_\nu(r)$ grows exponentially and the
function $K_\nu(r)$ decays exponentially for large positive $r$ \cite{r4}:
\begin{eqnarray}
I_\nu(r)&\sim&{1\over\sqrt{2\pi r}}\,e^r\qquad (r\to+\infty),\nonumber\\
K_\nu(r)&\sim&\sqrt{\pi\over2r}\,e^{-r}\qquad (r\to+\infty).
\label{eq2.19}
\end{eqnarray}
Eliminating the growing exponentials in Eqs.~(\ref{eq2.17}) and (\ref{eq2.18})
gives a pair of linear equations to be satisfied by the coefficients $C_1$
and $C_2$:
\begin{eqnarray}
C_1 e^{-\nu \pi i}+C_2 i\pi&=&0,\nonumber\\
C_1 e^{\nu \pi i}-C_2 i\pi&=&0.
\label{eq2.20}
\end{eqnarray}

A nontrivial solution to Eq.~(\ref{eq2.20}) exists only if the determinant
of the coefficients vanishes:
\begin{eqnarray}
{\rm det}\left(
\begin{array}{cc}
e^{-\nu \pi i}&i\pi\\
e^{\nu \pi i}&-i\pi
\end{array}\right)
=-2i\pi\cos(\nu\pi)=0.
\label{eq2.21}
\end{eqnarray}
\begin{table}
\caption[t1]{Comparison of the numerical values of $F(\epsilon)$ with that
predicted in Eq.~(\ref{eq2.22}) for the ground state $k=0$ of an
$x^2(ix)^\epsilon$ theory. The second column gives the exact values of the
ground-state energy for various values of $\epsilon$ in the first column.
In the third column is the value of $F$ obtained from the exact
energy in the second column using Eq.~(\ref{eq4.1}), which is a more precise
version of Eq.~(\ref{eq2.8}). Finally, in the fourth and fifth columns are the
first and second Richardson extrapolants \cite{r5} of the numbers in
the third column. Note that the exact values of $F(\epsilon)$ and their
Richardson extrapolants rapidly approach the asymptotic value $1/16=0.0625.$}
\begin{tabular}{ldddd}
$\epsilon$ & $E_0(\epsilon)$ & $F(\epsilon)$ & $R_1(\epsilon)$ & $R_2(\epsilon)$
\\ \tableline
8  & 5.55331  & 0.07825 &    -     &     -   \\
18 & 20.67629 & 0.06998 & 0.06336 &     -   \\
28 & 46.94324 & 0.06742 & 0.06281 & 0.06259\\
38 & 84.78728 & 0.06617 & 0.06266 & 0.06253\\
48 & 134.43752& 0.06542 & 0.06260 & 0.06251\\
58 & 196.03417 & 0.06493 & 0.06257 & 0.06251\\
\end{tabular}
\label{t1}
\end{table}
Hence, $\nu=k+{1\over2}$, and from Eq.~(\ref{eq2.11}), we have the {\it exact}
result
\begin{eqnarray}
F={1\over4}\left(k+{1\over2}\right)^2\qquad(k=0,1,2,3,\ldots).
\label{eq2.22}
\end{eqnarray}
Finally, we use Eq.~(\ref{eq2.8}) to obtain the large-$\epsilon$ behavior
of the eigenvalues $E$ given in Eq.~(\ref{eq1.5}). We verify this result
numerically in Table \ref{t1}.

\section{Arbitrary Integer $M$}
\label{s3}

The calculation of the energy levels for the general class of theories given in
Eq.~(\ref{eq1.6}) is a straightforward generalization of the calculation for the
case $M=1$ in Sec.~\ref{s2}. The crucial ingredient in the calculation is
understanding the array of Stokes' and anti-Stokes' lines along which we impose
the boundary conditions. This difference leads to $M$-dependent wave functions,
but the condition that determines the eigenvalues is still a simple
trigonometric equation.

Just as for the case $M=1$, we scale the differential equation (\ref{eq1.8})
using Eq.~(\ref{eq2.6}). In the limit as $\epsilon\to\infty$ the resulting
differential equation is identical to Eq.~(\ref{eq2.7}) except that now there is
a factor of $(-1)^{M+1}$ multiplying the exponential term. Again, we define $F$
as in Eq.~(\ref{eq2.8}) and change to the variable $w$ as prescribed by
Eq.~(\ref{eq2.9}). This gives the differential equation
\begin{eqnarray}
&& w^2{d^2\over dw^2}\psi(w)+w{d\over dw}\psi(w)\nonumber\\
&& \qquad-\left[(-1)^{M+1}w^2+\nu^2\right] \psi(w)=0,
\label{eq3.1}
\end{eqnarray}
which is the generalization of Eq.~(\ref{eq2.10}). In this equation $\nu$ is
defined as before by Eq.~(\ref{eq2.11}). Note that except for the appearance
of the $(-1)^{M+1}$ multiplying the $w^2$ term this equation 
is independent of $M$.

To solve Eq.~(\ref{eq3.1}) we consider the two cases of odd $M$ and even $M$
separately. If $M$ is odd this equation is identical to Eq.~(\ref{eq2.10}), and
the general solution is that given in Eq.~(\ref{eq2.12}). If $M$ is even
Eq.~(\ref{eq3.1}) is no longer a modified Bessel equation, but instead is just
the standard Bessel equation. Hence, in this case the general solution to
Eq.~(\ref{eq3.1}) is a linear combination of the ordinary Bessel functions 
$J_\nu$ and $Y_\nu$:
\begin{eqnarray}
\psi(w)=C_1J_\nu(w)+C_2Y_\nu(w).
\label{eq3.2}
\end{eqnarray} 
Thus, in terms of the variable $z$ the wave function in this case is
\begin{eqnarray}
\psi(z)=C_1J_\nu\left(\nu e^{i\pi z/2}\right)+C_2Y_\nu\left(\nu e^{i\pi z/2}
\right).
\label{eq3.3}
\end{eqnarray} 

Although it appears that this solution is independent of the parameter $M$, one
must recall that the boundary conditions do depend on $M$. Thus, the wave
functions and energy eigenvalues do indeed depend on $M$. To be precise, for a
given $M$ the Stokes' lines emanating from $z=\pm M$ are joined by a string of
adjacent arches of length $2$. The anti-Stokes' lines leave $\pm M$ and
asymptote to the lines ${\rm Re}\,z=\pm(M+1)$.  

For odd $M$ we impose the boundary conditions as for the case $M=1$ except that
the wave function $\psi$ vanishes along different lines. For even $M$ the wave
functions in Eq.~(\ref{eq3.3}) must first be expressed in terms of modified
Bessel functions using the functional equations \cite{r4}
\begin{eqnarray}
J_\nu(iz)&=&e^{\nu\pi i/2}I_\nu(z),\nonumber\\
Y_\nu(iz)&=&{-2 \over \pi} e^{-\nu\pi i/2}K_\nu(z)+ie^{\nu\pi i/2}I_\nu(z),
\label{eq3.4}
\end{eqnarray}
and then be treated using the same procedure as for odd $M$. 

Although the matrix elements for the linear equations obtained for odd $M$ and
even $M$ are quite different, the eigenvalue conditions are similar. For $M=2$
the condition is
\begin{eqnarray}
\cos(2\nu\pi)=-1/2,
\label{eq3.5}
\end{eqnarray} 
whose solution is
\begin{eqnarray}
\nu=k+{1\over3}\quad{\rm and}\quad\nu=k+{2\over3}.
\label{eq3.6}
\end{eqnarray}
Thus, for large $\epsilon$ the energy is given by
\begin{eqnarray}
E={1\over4}\left(k+{P\over3}\right)^2\epsilon^2,
\label{eq3.7}
\end{eqnarray}
where $P=1,2$. Note that this result is the $M=2$ case of Eq.~(\ref{eq1.7}).
This expression is verified numerically in Table \ref{t2} for the case of the
ground state energy corresponding to $k=0$ and $P=1$. For arbitrary $M$ one
obtains the result in Eq.~(\ref{eq1.7}).

\begin{table}
\caption[t2]{Comparison of the numerical values of $F(\epsilon)$ with that
predicted in Eq.~(\ref{eq3.7}) for the ground state $k=0,P=1$ of an
$x^4(ix)^\epsilon$ theory. The second column gives the exact values of the
ground-state energy for various values of $\epsilon$ in the first column.
In the third column is the value of $F$ obtained from the energy in the
second column using Eq.~(\ref{eq4.1}). Finally, in the fourth and fifth columns
are the first and second Richardson extrapolants \cite{r5} of the numbers in
the third column. Note that the exact values of $F(\epsilon)$ and their
Richardson extrapolants rapidly approach the asymptotic value
$1/36=0.0277778.$}
\begin{tabular}{ldddd}
$\epsilon$ & $E_0(\epsilon)$ & $F(\epsilon)$ & $R_1(\epsilon)$ & $R_2(\epsilon)$
\\ \tableline
8  & 2.65128 & 0.05035 &    -      &     -   \\
18 & 9.21477 & 0.03551 & 0.02661 &     -   \\
28 & 20.70525 & 0.03232 & 0.02722 & 0.02740\\
38 & 37.32010 & 0.03097 & 0.02746 & 0.02766\\
48 & 59.16865 & 0.03023 & 0.02756 & 0.02772\\
58 & 86.31766 & 0.02977 & 0.02764 & 0.02775\\
\end{tabular}
\label{t2}
\end{table}

Observe that the magnitude of the energy eigenvalues decreases as $M$ increases.
At first glance this might seem surprising, but it can be easily understood in
terms of the uncertainty principle. As $M$ increases, the anti-Stokes' lines on
which we impose the boundary conditions for the differential equation
(\ref{eq1.8}) move away from the negative imaginary axis, as we can see from
Eq.~(\ref{eq1.9}). For example, for fixed $\epsilon$ the anti-Stokes' lines for
$M=2$ are separated by a greater distance than for $M=1$; the anti-Stoke's lines
for $M=3$ are separated by a greater distance than for $M=2$, and so on. Hence,
the uncertainty in the position $\Delta x$ increases with $M$. By the
uncertainty principle, this increase in the uncertainty of the position
corresponds to a decrease in the uncertainty of the momentum, and thus, a 
decrease in the energy. This argument explains the large-$M$ behavior of the
result in Eq.~(\ref{eq1.7}).

\section{Higher-Order Corrections to the $\epsilon\to\infty$ Limit for $M=1$}
\label{s4}

In this section we show how to calculate the corrections to the large-$\epsilon$
behavior in Eq.~(\ref{eq1.5}). These corrections are of order $\epsilon$ and
$\epsilon\ln\epsilon$. From these higher-order calculations we obtain an
extremely accurate approximation $E_k$ for all $k$. Our asymptotic analysis
begins with the change of variable in Eq.~(\ref{eq2.6}), but we use a more
precise version of Eq.~(\ref{eq2.8}):
\begin{equation}
F(\epsilon)={E^{\epsilon+4\over\epsilon+2}\over(\epsilon+2)^2}.
\label{eq4.1}
\end{equation} 
We find that the function $F(\epsilon)$ is a series in inverse powers of
$\epsilon$ of the form $F=f_0+f_1\epsilon^{-1}+f_2\epsilon^{-2}+\cdots$. The
coefficient $f_0=F(\infty)$ is given in Eq.~(\ref{eq2.22}). Our objective here
is to calculate $f_1$, and from this to calculate the first correction to $E$.

In addition to $F(\epsilon)$, the wave function $\psi$ is also a series in
inverse powers of $\epsilon$, $\psi(z)=\psi_0+\psi_1(z)\epsilon^{-1}+\psi_2(z)
\epsilon^{-2}+\cdots$. Using this series and collecting like powers of
$\epsilon$ we obtain the following sequence of differential equations:
\begin{eqnarray}
&&\epsilon^0:\quad{d^2\over dz^2}\psi_0(z)+f_0\pi^2\left(1+e^{iz\pi}\right)
\psi_0(z)=0,\nonumber\\
&&\epsilon^{-1}:\quad{d^2\over dz^2}\psi_1(z)+f_0\pi^2\left(1+e^{iz\pi}\right)
\psi_1(z)\nonumber\\
&&\qquad\qquad=-\left[f_0\pi^2e^{iz\pi}{z^2\over 2}+f_1(1+e^{iz\pi})\right]\pi^2
\psi_0(z),\nonumber\\
&&\epsilon^{-2}:\quad{d^2\over dz^2}\psi_2(z)+f_0\pi^2\left(1+e^{iz\pi}\right)
\psi_2(z)\nonumber\\
&&\qquad\qquad=-\left[f_0\pi^2e^{iz\pi}{z^2\over2}+f_1(1+e^{iz\pi})\right]\pi^2
\psi_1(z)\nonumber\\
&&\qquad\qquad-\left[f_0e^{iz\pi}\left(z^2\pi^2+z^3\pi^3i-{z^4\pi^4\over8}
\right)
\right.\nonumber\\
&&\qquad\qquad\left.+f_1e^{iz\pi}{z^2\pi^2\over2}+f_2\left(1+e^{iz\pi}\right)
\right]\pi^2\psi_0(z).
\label{eq4.2}
\end{eqnarray}   

The first equation is exactly Eq.~(\ref{eq2.7}). The second equation contains
the coefficient $f_1$. To solve for $f_1$ we observe that the solution to the
homogeneous part of the equation is just the solution to the first equation.
This suggests using the method of reduction of order; to wit, we let $\psi_1(z)=
u_1(z)\psi_0(z)$. To solve the resulting equation for $f_1$ we then multiply by
$\psi_0(z)$ and integrate over the $WKB$ path with respect to $z$. The first
equation can be used to simplify this equation and the left side becomes the
expression $u_1(z)\psi_0^2(z)$ evaluated at the end points. Since $\psi_0(-i
\infty)=0$, the left side equals zero, and we can solve for $f_1$ in 
quadrature form. To be explicit,
\begin{eqnarray}
f_1=-{1\over2}f_0\pi^2{\int_{-2-i\infty}^{2-i\infty}dz\,z^2\psi_0^2(z)e^{iz\pi}
\over\int_{-2-i\infty}^{2-i\infty}dz\,\psi_0^2(z)(1+e^{iz\pi})}.
\label{eq4.3}
\end{eqnarray}

To prepare for evaluating these integrals we change to the variable $w$ in
Eq.~(\ref{eq2.9}) with $F=f_0$ and obtain
\begin{eqnarray}
f_1={1\over2}{\int_{-\infty+i\delta}^{-\infty-i\delta}dw\,w\psi_0^2(w)\ln^2
\left({w\over2\sqrt{f_0}}\right)\over\int_{-\infty+i\delta}^{-\infty-i\delta}
{dw\over w}\psi_0^2(w)\left(1+{w^2\over 4f_0}\right)}, 
\label{eq4.4}
\end{eqnarray} 
where $\delta$ is infinitesimal and the contour of integration
goes around the origin.

For the case $k=0$ these integrals are easy to evaluate because $f_0=1/16$ and 
$\psi_0(w)=I_{1/2}(w)+K_{1/2}(w)/\pi = e^w/\sqrt{2\pi w}$. Substituting these
expressions into Eq.~(\ref{eq4.4}) gives
\begin{eqnarray}
f_1={1\over2}{\int_{-\infty+i\delta}^{-\infty-i\delta}dw\,e^{2w}\ln^2(2w)\over
\int_{-\infty+i\delta}^{-\infty-i\delta}dw\,e^{2w}\left(4+{1\over w^2}\right)}.
\label{eq4.5}
\end{eqnarray}   
By carefully evaluating the discontinuities across the branch cut and the 
residues at the singularities of the integrands, we obtain
\begin{eqnarray}
f_1=\gamma/4,
\label{eq4.6}
\end{eqnarray}
where $\gamma$ is Euler's constant. Combining this result with Eq.~(\ref{eq4.1})
and solving for $E$ in the limit of large $\epsilon$ yields
\begin{eqnarray}
E={1\over16}\epsilon^2-{1\over4}\epsilon\ln\epsilon+{1\over4}(1+\gamma+2\ln2)
\epsilon+{\rm O}(\ln\epsilon).
\label{eq4.7}
\end{eqnarray}

By comparison, if we calculate to next order in WKB, we obtain for the $k$th
energy level
\begin{eqnarray}
E_k&\sim&\left[{\Gamma\left({8+3\epsilon\over4+2\epsilon}\right)\sqrt{\pi}(k+1/2
)\over\sin\left({\pi\over2+\epsilon}\right)\Gamma\left({3+\epsilon\over2+
\epsilon}\right)}\right]^{4+2\epsilon\over4+\epsilon}\nonumber\\
&\times& \left[1+{(2+\epsilon)(1+\epsilon)\sin\left({2\pi\over2+\epsilon}\right)
\over6\pi(k+1/2)^2(4+\epsilon)^2}\right] \quad(k\to\infty). 
\label{eq4.8}
\end{eqnarray} 
Taking the large $\epsilon$ limit of this expression gives
\begin{eqnarray}
E={\epsilon^2\over16}-{1\over4}\epsilon\ln\epsilon +{1\over4}\left({7\over3}+
\ln2\right)\epsilon + {\rm O}(\ln\epsilon). 
\label{eq4.9}
\end{eqnarray}
The appearance of a $\log\epsilon$ term in this behavior is a consequence of
the structure of Eq.~(\ref{eq4.1}). Note that the coefficent of the $\epsilon$
term for $WKB$ differs from the exact result but $WKB$ is numerically very
accurate. WKB gives $0.75662$ compared with $0.74088$ for the exact result.

In Table \ref{t3} the results of a Richardson extrapolation \cite{r5} of the
exact values of $F(\epsilon)$ are given. These results verify that the value of
$f_1$ is correct.

This work was supported in part by the U.S. Department of Energy.

\begin{table}
\caption[t1]{Comparison of the numerical value of the coefficient $f_1$ in
Eq.~(\ref{eq4.6}) with a fit to the exact values of $F(\epsilon)$ for the
case of the ground state $k=0$ of an $x^2(ix)^\epsilon$ theory. The second
column gives the exact values of the ground-state energy for various values of
$\epsilon$ in the first column. In the third column is the approximation to
$f_1$ obtained from $F(\epsilon)$ by subtracting off the leading
large-$\epsilon$ behavior given in Eq.~(\ref{eq2.22}). In the fourth and fifth
columns are the first and second Richardson extrapolants \cite{r5} of the
numbers in the third column. Note that the approximations in column 3 and their
Richardson extrapolants rapidly approach the asymptotic value 
of $f_1=\gamma/4=0.144304$.}
\begin{tabular}{ldddd}
$\epsilon$ & $E_0(\epsilon)$ & $R_0(\epsilon)$ & $R_1(\epsilon)$ & 
$R_2(\epsilon)$
\\ \tableline
8  & 5.55331  & 0.12597 &    -     &     -   \\
18 & 20.67629 & 0.13460 & 0.14150 &     -   \\
28 & 46.94324 & 0.13767 & 0.14321 & 0.14389\\
38 & 84.78728 & 0.13926 & 0.14372 & 0.14418\\
48 & 134.43752& 0.14024 & 0.14394 & 0.14425\\
58 & 196.03417& 0.14090 & 0.14406 & 0.14428\\
\end{tabular}
\label{t3}
\end{table}

\end{document}